\documentclass[aps, prd, twocolumn, amsmath, floats, floatfix, superscriptaddress, nofootinbib,
showpacs]{revtex4-2}

\usepackage{epsfig}
\usepackage{amsmath,amsfonts,amssymb}
\usepackage[caption=false, justification=centerlast]{subfig}
 
\usepackage{verbatim}
\usepackage{float}
\usepackage{morefloats}
\usepackage[dvipsnames]{xcolor}
\usepackage[normalem]{ulem}
\usepackage{graphicx}

\usepackage{multirow}
\usepackage{makecell}
\usepackage{tabularx}
\usepackage{cancel}

\usepackage{url}
\usepackage{breakurl}
\usepackage[breaklinks]{hyperref}

\usepackage[T1]{fontenc}

\usepackage[capitalise]{cleveref}

\crefname{section}{Sec.}{Secs.}
\crefname{table}{Tab.}{Tabs.}
\crefname{figure}{Fig.}{Figs.}
\crefname{equation}{Eq.}{Eqs.}
\crefname{appendix}{Appendix\ }{Appendix\ }

\providecommand{\openone}{\leavevmode\hbox{\small1\kern-3.8pt\normalsize1}}

\usepackage[Q=yes,pverb-linebreak=no]{examplep}
\usepackage{aas_macros}
\usepackage{scrextend}

\usepackage{siunitx}

\usepackage{nameref}

\begin{document}

\title{\textbf{Identification of Binary Neutron Star Mergers in Gravitational-Wave Data\\ Using YOLO One-Shot Object Detection}}

\author{Jo\~ao Aveiro}
\email{joao@aveiro.me}
\affiliation{CFisUC, 
	Department of Physics, University of Coimbra, 3004-516  Coimbra, Portugal}

\author{Felipe~F.~Freitas}
\email{felipefreitas@ua.pt}
\affiliation{Departamento de F\'\i sica da Universidade 
de Aveiro and \\
Centre  for  Research  and  Development  in  Mathematics  and  Applications  (CIDMA)\\
Campus de Santiago, 3810-183 Aveiro, Portugal}

\author{Márcio Ferreira}
\email{marcio.ferreira@uc.pt}
\affiliation{CFisUC, 
	Department of Physics, University of Coimbra, 3004-516  Coimbra, Portugal}

\author{Antonio~Onofre}
\email{antonio.onofre@cern.ch}
\affiliation{Centro de F\'{\i}sica das Universidades do Minho e do Porto (CF-UM-UP), Universidade do Minho, 4710-057 Braga, Portugal}

\author{Constança Providência}
\email{cp@uc.pt}
\affiliation{CFisUC, 
	Department of Physics, University of Coimbra, 3004-516  Coimbra, Portugal}

\author{Gon\c{c}alo Gon\c{c}alves}
\email{goncalo.mota.goncalves@gmail.com}
\affiliation{LIP, Department of Physics, University of Coimbra, 3004-516 Coimbra, Portugal}

\author{José A. Font}
\email{j.antonio.font@uv.es}
\affiliation{Departamento de Astronomía y Astrofísica, Universitat de València,
Dr. Moliner 50, 46100, Burjassot (València), Spain}
\affiliation{Observatori Astronòmic, Universitat de València,
Catedrático José Beltrán 2, 46980, Paterna (València), Spain}

\begin{abstract}
    We demonstrate the application of the YOLOv5 model, a general purpose convolution-based single-shot object detection model, in the task of detecting binary neutron star (BNS) coalescence events from gravitational-wave data of current generation interferometer detectors. We also present a thorough explanation of the synthetic data generation and preparation tasks based on approximant waveform models used for the model training, validation and testing steps. Using this approach, we achieve mean average precision (mAP\textsubscript{[0.50]}) values of 0.945 for a single class validation dataset and as high as 0.978 for test datasets. Moreover, the trained model is successful in identifying the GW170817 event in the LIGO H1 detector data. The identification of this event is also possible for the LIGO L1 detector data with an additional pre-processing step, without the need of removing the large glitch in the final stages of the inspiral. The detection of the GW190425 event is less successful, which attests to performance degradation  with the signal-to-noise ratio. Our study indicates that the YOLOv5 model is an
    interesting approach for  first-stage detection alarm
    pipelines and, when integrated in more complex pipelines, for real-time inference of physical source parameters.
\end{abstract}

\maketitle
 
\section{Introduction}

Matched-filtering has been a highly successful technique used in gravitational-wave (GW) astronomy for the search of compact binary coalescences \cite{Owen:1998dk}. The method relies on waveform templates that cover a large variety of configurations for the binary parameter space. These templates are constructed using inspiral-merger waveform models (see \cite{Dietrich:2018uni} for a discussion on the different BNS waveform approximants). The matched-filtering technique has allowed the detection of multiple compact binary coalescences by the Advanced LIGO and Advanced Virgo GW observatories~\cite{abbott2021gwtc1,abbott2021gwtc2,abbott2021gwtc3}. A considerable increase in the number of observed events is expected from future runs of the LIGO \cite{LIGOScientific:2014pky}, Virgo \cite{VIRGO:2014yos}, and KAGRA \cite{PhysRevD.88.043007} observatories~\cite{Prospects}. The detection of possible electromagnetic counterparts from compact binary coalescences requires real-time identification of potential candidates and rapid inference of the source location. These requirements contrast with the expensive computational demand and poor scaling properties of matched-filtering techniques. Besides, data recorded by GW detector is affected by non-stationary noise transients \cite{Tiwari:2015ofa,LIGOScientific:2018kdd}, known as glitches, which adds a complexity layer to data analysis pipelines. 

Machine learning (ML) is becoming a powerful approach in several problems of GW physics where the list of applications of neural networks (NN) is rapidly growing (see~\cite{Cuoco:2020ogp} and references therein).  
Deep learning (DL) has been explored as an alternative and/or complement to  matched-filtering. The different DL models in the literature can generally be divided into classification models, which try to distinguish GW signals from noise, or regression models, where multiple-parameter estimation is the goal.  Convolutional Neural Networks (CNNs) have been applied for GW searches, already equaling the accuracy of the matched-filtering-based approach in binary black hole (BBH) mergers~\cite{Gabbard:2017lja,George:2016hay,Schafer:2021fea}. The fast inference of CNNs makes them natural candidates for trigger generators \cite{Gebhard:2019ldz}. For BNS mergers, a Deep Neural Network (DNN) was employed in \cite{Dreissigacker:2019edy} to distinguish noise from signal directly from the detector strain data. A sensing layer was added to a CNN to improve weak signal recognition in \cite{Wang:2019zaj}. A NN-based algorithm to detect signals from BNS mergers in the detectors’ strain data was introduced in \cite{Schafer:2020kor}. CNNs have also been explored to identify and distinguish BNS from BBH signals and noise \cite{Krastev:2019koe,Krastev:2020skk,Menendez-Vazquez:2020khz}. Batch normalization and dropout methods were applied on a CNN model to detect BBH signals in \cite{Xia:2020vem}. In \cite{Lin:2020aps}, a combination of CNNs, Long Short-Term Memory Recurrent Neural Networks, and DNNs was explored to detect BBH events. More recently, an ensemble algorithm of CNNs models for GW signal recognition was analyzed in \cite{Ma:2022esx}. Even without significant optimizations of the CNNs, their use in~\cite{Alvares:2020bjg}, showed that results obtained with DL algorithms were consistent to the large majority of the results published by the LIGO-Virgo Collaborations.

The approach we put forward in this paper contrasts with previous works in that we propose using general-purpose object detection models used in Computer Vision (CV) tasks. These models generally provide precise time bounds for any detected event, whilst being compatible with real-time operation with reduced computational resources \cite{Sultana_2020, 8627998} in single detector applications. Additionally, object detection models are usually geared towards multi-class detection, thus allowing for future development of detection pipelines that provide simultaneous detection of various event types - e.g. all families of compact binary coalescence events, glitches, etc. Here, we demonstrate the suitability of the YOLOv5 model, a general purpose convolution-based single-shot object detection model, to detect GW signals from BNS mergers using data from current generation interferometer detectors.

The paper is organized as follows: A  thorough description of the proposed approach and of the object detection model  is presented in Section \ref{sec:model_description}. The generation of the dataset used for this task, including the simulation of GW time-series and their consequent conversion to spectrograms, is described in Section \ref{sec:gen}. The training step of the proposed model is presented in Section \ref{sec:train}. In Section \ref{sec:tests} we analyze the effects of varying the signal-to-noise (SNR) ratio of the GW injections as well as the impact of different object/background ratios on the performance of the model, and additionally, we provide real-world test results for the detection of the two BNS merger events reported by the LIGO-Virgo Collaboration, GW170817 and GW190425. Lastly, our conclusions are drawn in Section \ref{sec:conclusions}.

\section{Proposed approach and model description} \label{sec:model_description}

Object detection is usually described in the field of computer vision (CV) as the task of identifying the presence and the location of specific instances of a real-world object class in an image. This task is of central importance in CV and has broad applicability in various fields. As such, a wide variety of highly mature and diverse models and methodologies have been developed in recent years \cite{https://doi.org/10.48550/arxiv.1905.05055, Sultana_2020, 8627998}. Due to the diversity of target hardware for real-time inference, these models are usually extremely efficient to be able to operate even in low-powered devices.

In this paper we propose the application of a general-purpose object detection model for the task of detecting GW signals from BNS mergers in the raw interferometer data from current generation detectors. The data resulting from such detectors, being a uni-dimensional time series array of numerical data, does not contemplate a conventional sample capable of being provided to such a model. Notwithstanding, we propose the usage of spectrogram images as the data source of our model.

This approach, whilst not the most orthodox, allows for taking advantage of the maturity and performance of such models. More so, with implementations that allow for multi-class detection, event classification tasks can also be addressed - e.g. in  glitch detection and classification. Finally, these models can be implemented in low-cost, low-power equipment infrastructures and provide a real-time, low-latency detection pipeline for candidate event alarms.

We utilize a model from the You Only Look Once (YOLO) object detection model family \cite{https://doi.org/10.48550/arxiv.1506.02640, https://doi.org/10.48550/arxiv.1612.08242, https://doi.org/10.48550/arxiv.1804.02767, https://doi.org/10.48550/arxiv.2004.10934, glenn_jocher_2022_6222936} for detecting GW signals from coalescing BNS events. The YOLO family consists of convolution-based single-shot object detection models that have been proven to provide good detection capabilities whilst remaining extremely lightweight \cite{Sultana_2020, 8627998}, allowing for real-time inference tasks on hardware with low computational resources.
Specifically, we utilize the Ultralytics YOLOv5 library \cite{glenn_jocher_2022_6222936}, which implements a version of these models based on the PyTorch framework \cite{NEURIPS2019_9015}. Apart from providing various model architectures, varying in parameter size and complexity, the Ultralytics implementation also offers extensive functionality for training, testing, and deploying such models, which allows for fast and efficient development of new detection solutions for a wide range of tasks. Additionally, various tools for converting the trained models to other architectures - such as TensorRT, TFLite, and ONNX - are provided, thus aiding the deployment of performant,  production-ready inference pipelines.

The ease of usage of the YOLOv5 implementation, as well as its capabilities and performance, make it the ideal candidate for quickly and efficiently developing the proof of concept proposed in this work. Nevertheless, various other object detection models and implementations are widely available and should be considered for future analysis - different versions or variants of the YOLO family, such as YOLOv4 \cite{https://doi.org/10.48550/arxiv.2004.10934}, YOLOR \cite{https://doi.org/10.48550/arxiv.2105.04206}, and YOLOX \cite{https://doi.org/10.48550/arxiv.2107.08430}, and other competitive solutions, such as SSD \cite{Liu_2016}, RefineDet \cite{https://doi.org/10.48550/arxiv.1711.06897}, Fast R-CNN \cite{https://doi.org/10.48550/arxiv.1504.08083}, and Faster R-CNN \cite{https://doi.org/10.48550/arxiv.1506.01497}, amongst others.

\section{Data generation and processing} \label{sec:gen}

\subsection{Waveform generation}

In this paper we use the \texttt{PyCBC} package \cite{alex_nitz_2022_6324278} for generating the GW signals used in the training, validation, and testing steps of the YOLOv5 model. Our approach is loosely based on the generation pipeline proposed by  \cite{Gebhard_2019, timothy_gebhard_2019_2649359}.

GW strain data from compact binary coalescences were generated in the time domain using the IMRPhenomPv2-NRTidalv2  waveform approximant \cite{Dietrich_2019}. The values of the component masses - $M_1$ and $M_2$ - were sampled using a random number generator (RNG) considering a uniform distribution in $M \in [1.0, 2.0[ \si{M_\odot }$, while the values of the  dimensionless tidal deformability parameters - $\Lambda_1$ and $\Lambda_2$ - were sampled randomly in $\Lambda \in [0.0, 1000.0 [$, respecting the condition $M_1 > M_2 \Rightarrow \Lambda_1 < \Lambda_2$. The inclination was sampled from a sine distribution and in all instances the individual spins were set to zero.

The generated detector-agnostic waveforms are projected onto the antenna pattern of the LIGO H1 detector. The physical parameters of the projection are sampled from distributions that provide the most uniform coverage in the parameter space, namely right ascension and declination distributions for uniform sky coverage, and a uniform distribution for the polarization angle.

To better simulate the operation of the detector, we must also consider a noise model. To do so, we inject the waveforms in detector Gaussian noise based on the noise power spectral density (PSD) of the Advanced LIGO detectors, using routines provided by \texttt{PyCBC}\footnote{Namely, we use 
the pycbc.psd.analytical.aLIGOZeroDetHighPower PSD and the pycbc.noise.noise\_from\_psd() function.}.
For this task, we first generate a noise sample with duration $t_N=\SI{40}{s}$. Then, the merger waveform is generated, sliced to half the duration of the noise sample - i.e. $t_M = \SI{20}{s}$ -, preserving the last stages of the event, and injected in the first half of the noise sample. Consequently, the merger event - which coincides with the time at which the maximum amplitude of the strain is attained - occurs approximately in the middle of the resulting waveform sample. 	

For the injection, the event SNR \footnote{The Signal-to-Noise ratio (SNR) is generally defined by $\text{SNR}=P_{\text{signal}}/P_{\text{noise}}$, where $P_{\text{signal}}$ and $P_{\text{noise}}$ is the average power of the signal and the noise, respectively.} is calculated and the coalescing waveform is scaled to a target SNR. The value of the target SNR is sampled uniformly in the [10, 20] range, assuming integer values.

After merging both signals, a whitening filter with a segment duration $t_{\rm seg}=\SI{4}{s}$ and maximum duration $t_{\rm filter}=\SI{2}{s}$ is applied. After whitening, the resulting samples are cut symmetrically to the desired duration of $\SI{16}{s}$ and saved in an HDF5 archive with 32-bit floating-point precision. For each sample, a variety of other parameters are also saved, namely the physical parameters of the binary, and the target SNR used. Both the noise and the strain are generated using a sample rate of $f=\SI{4096}{Hz}$.

\subsection{Spectrogram generation}

\begin{figure}[!ht]%
    \centering
    \includegraphics[width=\linewidth]{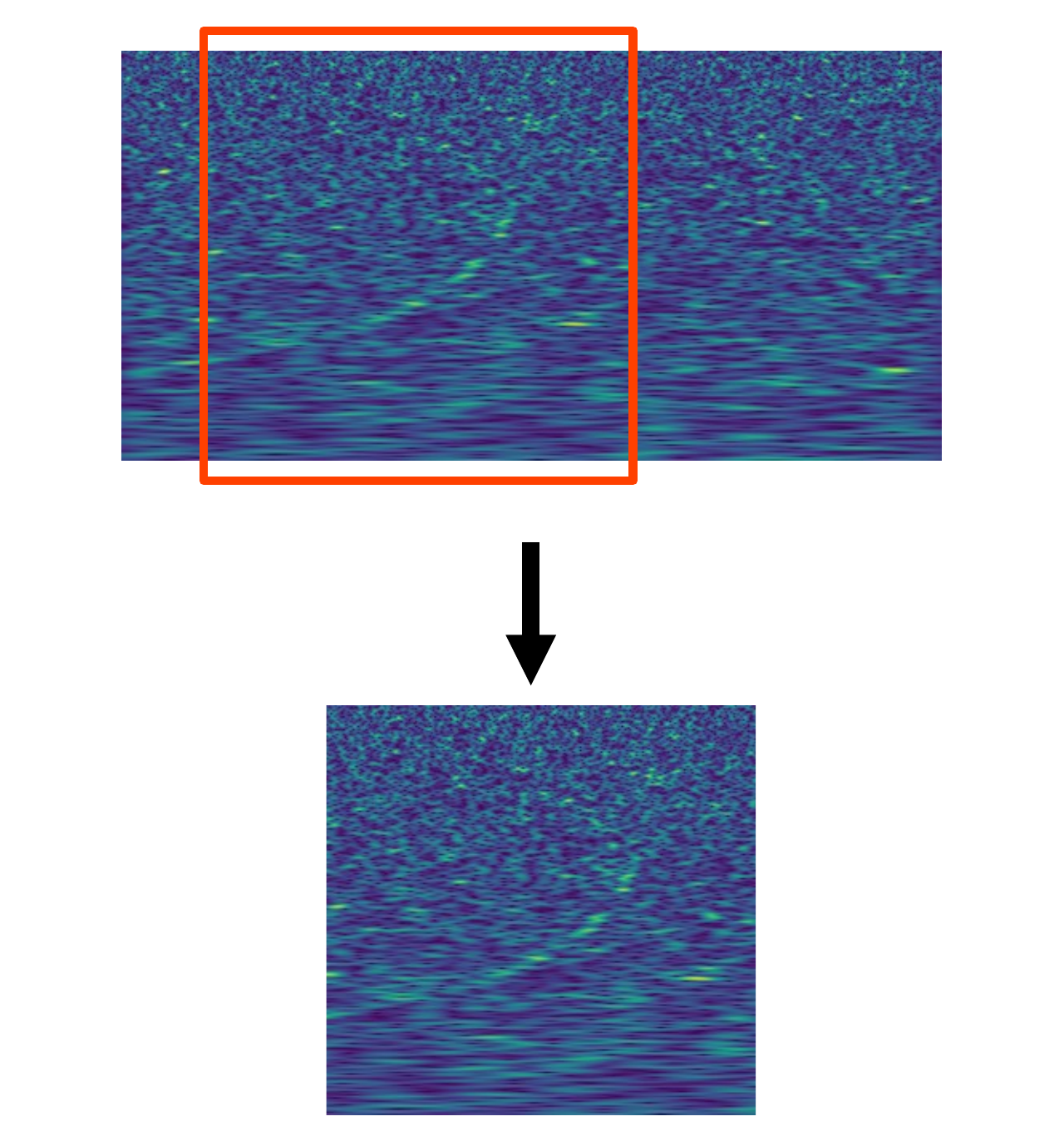}%
    \caption{Example of a generated spectrogram with \SI{16}{s} of duration (top) and final $\SI{8}{s}$ sample obtained by a random cut (bottom). The injection is placed on the first half of the original spectrogram and the random cut is done with the constraint that at least half of the final spectrogram sample includes the injection signal.}%
    \label{fig:example_cut}%
\end{figure}

After its generation, each waveform sample is cut to the final duration of $\SI{8}{s}$. This cut is performed randomly, with the constraint that any injection, if present, is preserved for at least $\SI{4}{s}$ - half of the final sample duration - cf. Figure \ref{fig:example_cut}. This allows for more variability in the relative position of the injected signals in the samples.

For converting the generated waveform samples into spectrograms, we use the Constant-Q Transform algorithm \cite{schorkhuber_2010} provided by the \texttt{librosa} package \cite{brian_mcfee_2022_6097378}. The transform is applied considering a minimum frequency of $f_{\rm min}=\SI{32}{Hz}$, with $n_{\text{bins}}=256$ frequency bins, $n_{\text{oct}}=64$ bins per octave and the tuning parameter set to 0, thus resulting in a maximum spectrogram frequency of $f_{\rm max} = \SI{512}{Hz}$. The hop length chosen is $n_{\rm hop} = 128$, which leads to 257 bins in the time axis for a sample with $\SI{8}{s}$ of duration. 

Each spectrogram is then converted to a color mesh (8-bit RGB values) and saved as a lossless PNG image file, with the origin of the time and frequency axis in the lower-left corner. The \texttt{viridis} colormap provided by the \texttt{matplotlib} library is used, and an automatic normalization of the color scale is performed in the color conversion process.

Each resulting image has a resolution of $257\times256 \si{px}$.  This resolution is the result of the parameters described previously for the Constant-Q Transform routine and no interpolation process is applied.

\subsection{Labeling}

Each image is accompanied by a text file including information regarding the boxes that define the objects present. For each object, the information includes the class, the height, width, and position of the respective bounding box. In this dataset, only one class is considered, referring to a general GW event. Also, only one object - i.e. one GW event - is contained within each sample, so each text file only describes one bounding box, in the case a GW strain is present, or it is empty, in the case it only contains noise.

\begin{figure}[!t]%
    \centering
    \includegraphics[width=\linewidth]{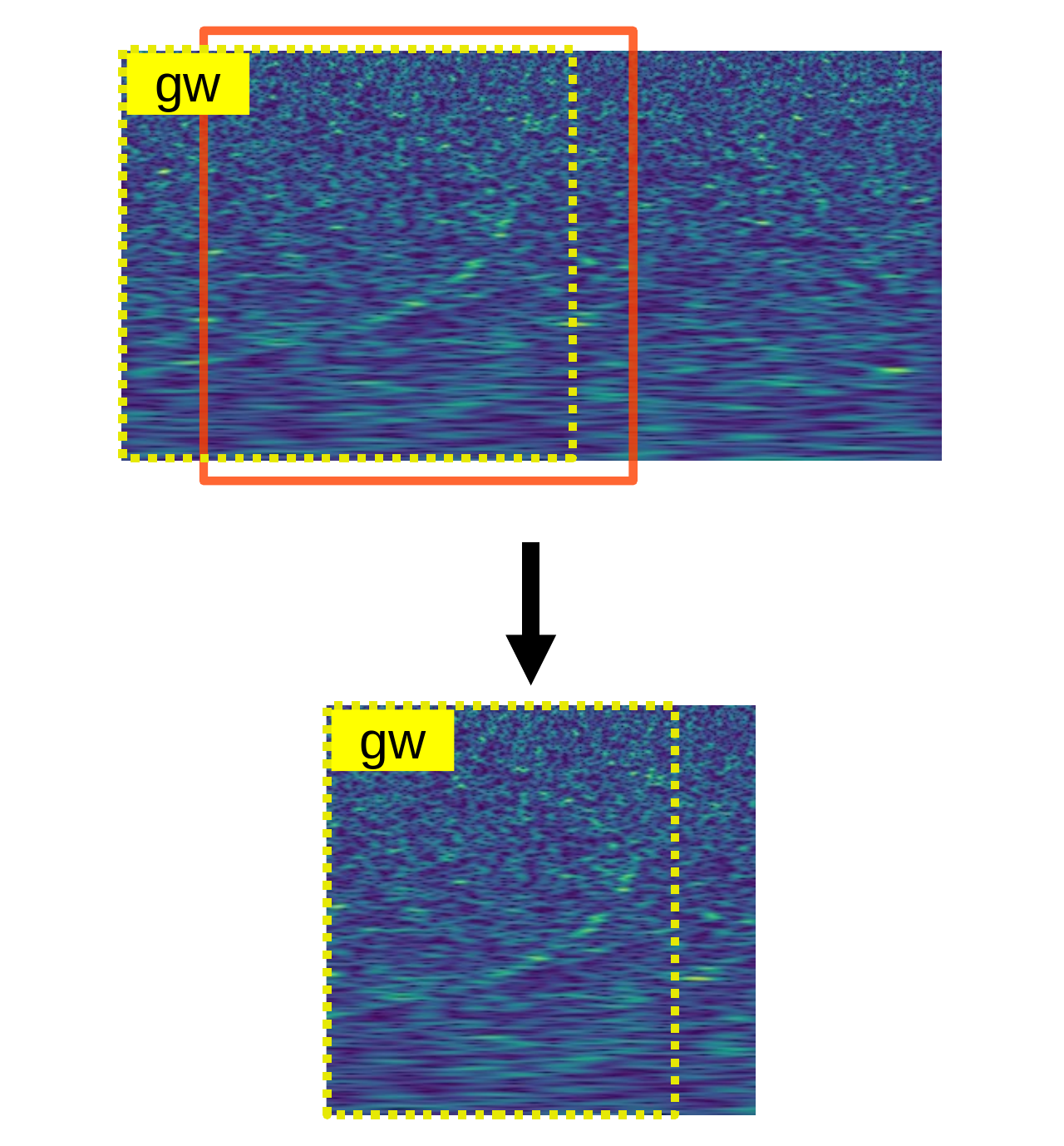}%
    \caption{Automatic labeling of the injection object. Since all injections occur in the first half of the uncut waveform samples, the bounding box definition (yellow box) of the cut samples can be obtained by the intersection of the cut window and the original GW strain bounding box.}%
    \label{fig:example_label}%
\end{figure}

The labeling of each object is done automatically alongside the data generation process - cf. Figure \ref{fig:example_label}. The bounding box width for each object is obtained from the time slice containing the injection. Since it is difficult to define a spectral region in which the injection occurs, the height of the bounding box is set to the height of the image, thus corresponding to the full frequency range considered for the transform.

\subsection{Dataset}

For training and validation tasks, a dataset with $10^5$ images was generated. The generated samples were randomly split into training and validation sets with ratios $n_{\rm train}/n_{\rm total}=\SI{80}{\%}$ and $n_{\rm val}/n_{\rm total}=\SI{20}{\%}$. For this task, we not only generated samples containing GW events - \textit{injections} - but also background only samples - purely noise from the detector. Although not required for the training of our model, the inclusion of background samples should provide better real-life results, namely reducing false positive detections.

The generated dataset consists of $\SI{50}{\%}$ of samples with injections and $\SI{50}{\%}$ of background samples. Considering the training/validation split is random, it is expected that each of the resulting subsets to be comprised of roughly $\SI{50}{\%}$ of injection samples and $\SI{50}{\%}$ of background samples. The actual dataset composition is described in Table \ref{tab:dataset_size}.

\begin{table}[!ht]
\centering
\caption{Composition of training and validation datasets. The values correspond to the number of samples of each type present in these subsets. Object samples correspond to spectrograms containing an event, in opposition to background samples which only contain simulated detector noise.}
\label{tab:dataset_size}
\begin{ruledtabular}\begin{tabular}{lll}
\textbf{Subset}    & \textbf{Background} & \textbf{Object}\\ 
Training       &   39923   &   40077        \\
Validation  &   10077   &   9923     \\
\end{tabular}\end{ruledtabular}%
\end{table}

\section{Model training} \label{sec:train}

The base model used in this work is the ``Large`` variant of the YOLOv5 v6 models - YOLOv5L6 - with an input size of 256x256 pixels. 

For the training, we relied on a transfer learning approach, having started with the pre-trained weights using the COCO 2017 Train dataset \cite{https://doi.org/10.48550/arxiv.1405.0312} provided by the Ultralytics YOLOv5 project. The optimizer used was the Stochastic Gradient Descent (SGD) with an initial learning rate of $\text{LR}_0=2.0 \times 10^{-6}$. The batch size was determined automatically to optimize the available graphical memory, resulting in a value of $154$. Other hyper-parameters and configurations were left unchanged from the default values provided by the Ultralytics YOLOv5 library.

The training was run for 200 epochs, having obtained converging loss and mean average precision (mAP) values. The best mAP value was achieved in the last epoch of training. The metrics obtained for this epoch are reported in Table~\ref{tab:train_metrics}. The evolution of the metrics and losses over the training epochs is provided in Appendix \ref{appendix:metrics}.

\begin{table}[!ht]
\centering
\caption{Precision, Recall, Mean Average Precision [0.50] (mAP\textsubscript{0.50}), and Mean Average Precision [0.50:0.95] (mAP\textsubscript{0.50:0.95}) values of the model training for the best epoch (200).}%
\label{tab:train_metrics}
\begin{ruledtabular}\begin{tabular}{llll}
\textbf{Precision} & \textbf{Recall} & \textbf{mAP\textsubscript{0.5}} & \textbf{mAP\textsubscript{0.5:0.95}} \\ 
0.922& 0.823 & 0.945     & 0.893        \\
\end{tabular}\end{ruledtabular}%
\end{table}

The instance used for training consisted of 32 vCPU cores, 64 GB of RAM, and a single NVIDIA V100 PCIe 16GB GPU. Using this setup, the total training time - including data loading and caching - was close to 14 hours for the dataset with 100 000 samples (80 000 for training and 20 000 for validation) of size $257\times256\si{px}$.

\section{Tests} \label{sec:tests}

\subsection{Unbalanced test sets}

As a first test of the validity of the model trained, three test datasets were generated using various ratios of background/injection samples, namely $20/80$, $50/50$, and $80/20$. Apart from this ratio, the dataset generation parameters follow the description provided above and are identical to the training and validation sets used, namely considering the same physical priors and a random SNR uniformly sampled in the [10, 20] interval. Each of these datasets is comprised of 20 000 samples. 

An inference pipeline was run for each dataset using a confidence threshold of 0.001 and 0.6 as the intersection-over-union (IoU)~\footnote{The intersection-over-union (IoU) is defined by the ratio between the area of overlap and the area of union between the detection and the ground-truth bounding boxes} threshold. The inference metrics for each dataset are presented in Table \ref{tab:test_unb}.

\begin{table}[!ht]
\centering
\caption{Precision, Recall, Mean Average Precision [0.50] (mAP\textsubscript{0.50}), and Mean Average Precision [0.50:0.95] (mAP\textsubscript{0.50:0.95}) values of test datasets with ratios 20/80, 50/50, and 80/20 (\textbf{\%\textsubscript{obj}/\%\textsubscript{bg}}) of object/background samples.}%
\label{tab:test_unb}
\begin{ruledtabular}\begin{tabular}{lllll}
\textbf{\%\textsubscript{obj}/\%\textsubscript{bg}}    & \textbf{Precision} & \textbf{Recall} & \textbf{mAP\textsubscript{0.5}} & \textbf{mAP\textsubscript{0.5:0.95}} \\ 
20/80 & 0.916     & 0.771  & 0.890     & 0.855         \\
50/50 & 0.933     & 0.820   & 0.947    & 0.894         \\
80/20 & 0.931    & 0.902  & 0.978    & 0.914         \\ 
\end{tabular}\end{ruledtabular}%
\end{table}

The test results show a degradation of the performance of the model - notorious in the recall and mAP metrics - possibly correlated with the increase of background-only samples in the datasets. Despite this, the overall metrics show promising results, and the degradation observed can possibly be improved by using larger datasets with more background samples, introducing real noise with greater variability, and/or implementing data augmentation techniques

\subsection{Signal-to-noise ratio metrics}

It is of interest to analyze the influence of the SNR in the model test metrics. As such, 11 datasets with fixed SNR in the range $[10, 20]$ were generated. Each dataset is comprised of 5000 samples with an injection/background ratio of $n_{\rm inj}/n_{\rm bg} \rightarrow 0.5/0.5$. The physical priors are the same as described previously. The confidence and IoU thresholds are the same as for the previous test.

\begin{figure*}[!t]%
    \centering
    \subfloat[\centering mAP\textsubscript{0.50}]{
        \includegraphics[width=0.45\linewidth]{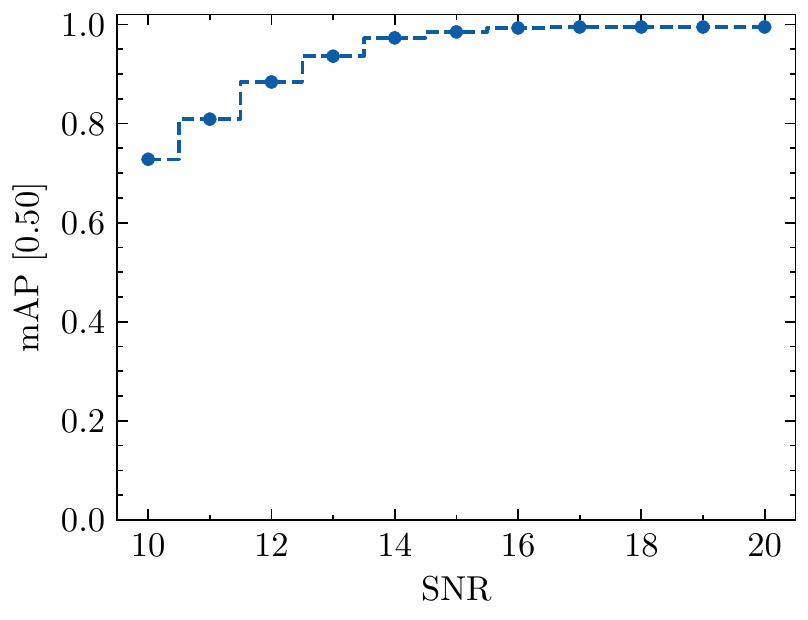} 
    }%
    \hfill
    \subfloat[\centering mAP\textsubscript{0.50:0.95}]{
        \includegraphics[width=0.45\linewidth]{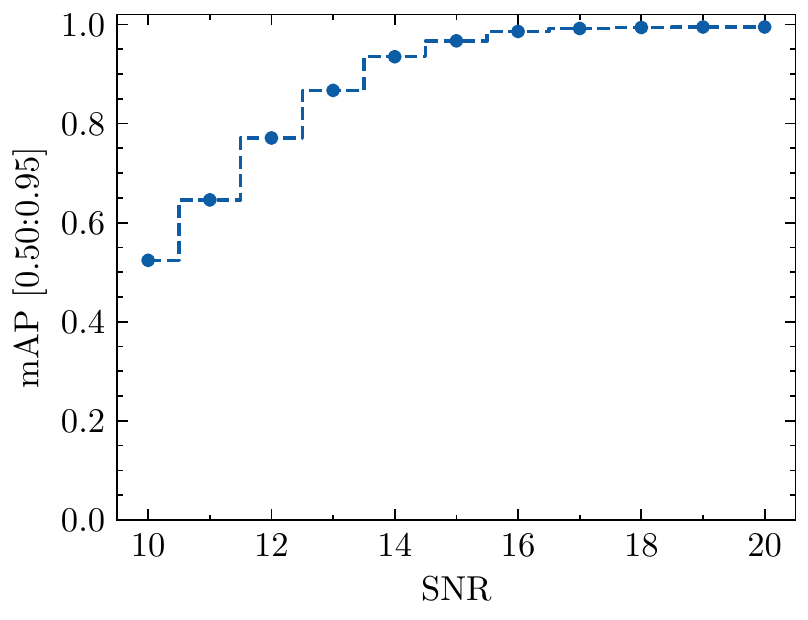} 
    }%
    \newline
    \medskip
    \centering
    \subfloat[\centering Precision]{
        \includegraphics[width=0.45\linewidth]{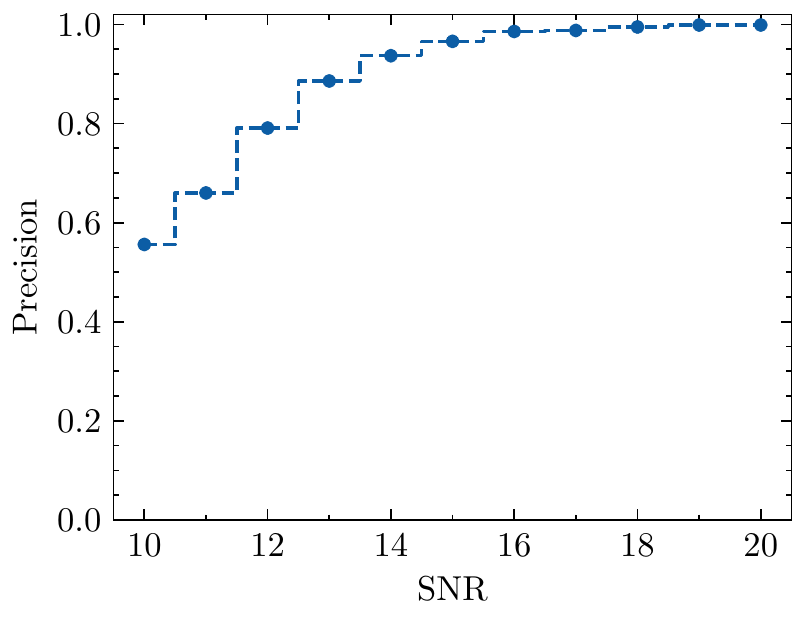} 
    }%
    \hfill
    \subfloat[\centering Recall]{
        \includegraphics[width=0.45\linewidth]{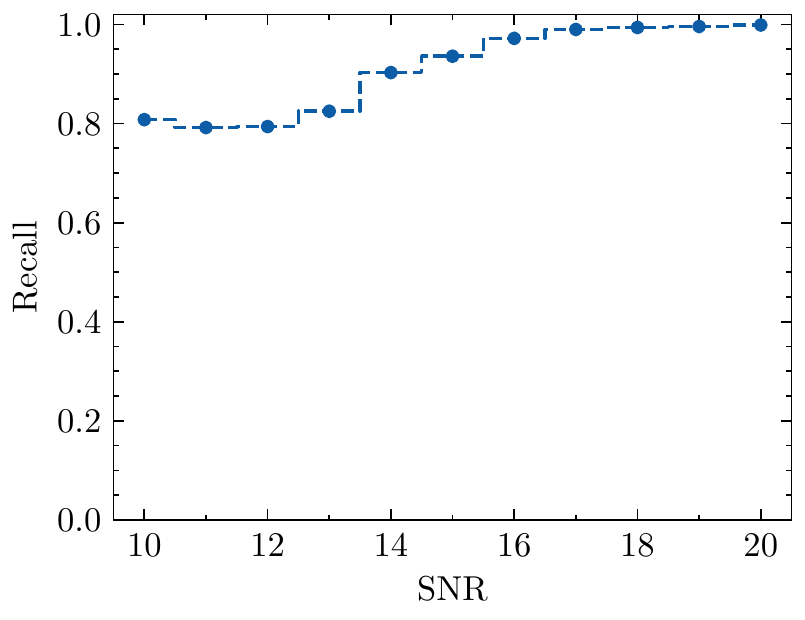} 
    }%
    
    \caption{Mean Average Precision [0.50] (mAP\textsubscript{0.50}), Mean Average Precision [0.50:0.95] (mAP\textsubscript{0.50:0.95}), Precision, and Recall values for different test datasets with SNR $\in [10, \,20]$ comprised of 5000 samples each.} %
    \label{fig:test_snr}%
    \medskip
\end{figure*}

The variation of the mAP\textsubscript{0.50}, mAP\textsubscript{0.50:0.95}, precision, and recall metrics as a function of the test dataset SNR are displayed in Figure \ref{fig:test_snr}.
The performance of the model is shown to improve greatly for larger (>15) SNR events, achieving nearly perfect metrics results. However, for samples with lower SNR, there is a significant degradation in all the metrics presented. In real-world tasks, this could imply a greater occurrence of false positives and/or false negatives if the confidence threshold is reduced in an attempt to detect such events. It can also be seen in Figure \ref{fig:test_snr} that the degradation of the precision metric is more severe than that of the recall metric, which could imply an increase of mostly false-positive detections.

The worse performance for lower SNR samples could imply the inability of the model architecture to learn the features and detect such events. Even so, it is probable that this effect arises from the bit-depth resolution limit resulting from the usage of 8-bit RGB pseudo-color images for the model input. For lower SNR, the pixel resolution provided by the data structure might not be sufficient to correctly preserve the details of such an event. This effect could be worsened by the usage of a pseudo-color map, as the available information symbols - ideally 24 bits (three 8-bit integer channels) - are effectively reduced.

\subsection{Detecting the GW170817 event}

Despite not being trained using real Compact Binary Coalescences (CBC) detections and real detector noise, it is of interest to analyze the ability of the trained model to perform inference in real event data. For this, we first consider the data of the GW170817 BNS coalescence event  \cite{Abbott_2019} obtained from the LIGO Hanford (H1) and the LIGO Livingston (L1) observatories, with a combined SNR of 32.4 \cite{PhysRevLett.119.161101}. 

For this test, a window of 602 seconds centered at the time of the event $t_e$ - i.e. the merger instant - was considered. A whitening filter was applied and the resulting corrupted samples were removed, reducing the overall duration of the data sample to $\SI{600}{s}$. The whitened waveform was then split using a window with a duration of $\SI{8}{s}$ and a stride of $\SI{4}{s}$, thus resulting in a total of 150 samples. For each sample, a spectrogram image was obtained by applying the transform described previously. The resulting images were provided to an inference pipeline with the trained weights. For this task, the Intersection over Union (IoU) threshold was set to $0.45$ and the confidence threshold was set to $0.90$.

\begin{figure}[!ht]%
    \centering
    \subfloat[\centering $t_e-8s<t <t_e$]{{\includegraphics[width=0.45\linewidth]{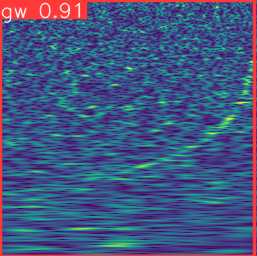} }}%
    \hfill
    \subfloat[\centering $t_e-4s<t <t_e+4s$]{{\includegraphics[width=0.45\linewidth]{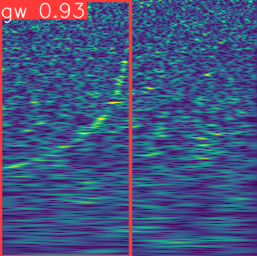} }}%
    \caption{Spectrograms of the GW170817 event data from the LIGO H1 detector and respective event detection bounding boxes and confidence value obtained by applying the trained YOLOv5 model}%
    \label{fig:h1v3}%
\end{figure}

\begin{figure}[!ht]%
    \centering
    \subfloat[\centering $t_e-8s<t <t_e$]{{\includegraphics[width=0.45\linewidth]{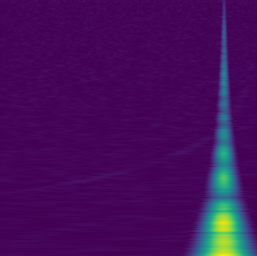} }}%
    \hfill
    \subfloat[\centering $t_e-4s<t <t_e+4s$]{{\includegraphics[width=0.45\linewidth]{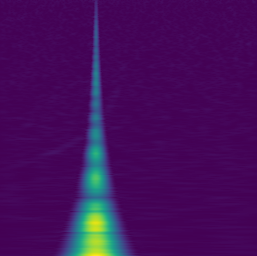} }}%
    \caption{Spectrograms of the GW170817 event data from the LIGO L1 detector with visible glitch}%
    \label{fig:l1v3}%
\end{figure}

For the H1 detector data, the event was correctly detected in the frames $t_e-8s<t <t_e$ and  $t_e-4s<t <t_e+4s$, see Figure \ref{fig:h1v3}. No false positives were detected.
For the L1 detector data, no detection compatible with the provided threshold - either a false positive or a true positive - was achieved. It should be noted, however, that the data from this detector suffers from an extensive glitch at the end of the inspiral. Due to the large amplitude of the glitch, the spectrogram normalization leads to an image that is visually vastly different from the samples provided in the model training step, as can be seen in Figure \ref{fig:l1v3}. 

\begin{table}[!ht]
\centering
\caption{Minimum ($\min{x}_{ij}$), maximum ($\max{x}_{ij}$), mean ($\overline{x}_{ij}$), standard deviation ($\sigma_{{x}_{ij}}$), and median ($\widetilde{x}_{ij}$) of spectrogram values for the $t_e-4s<t <t_e+4s$ time interval GW170817 data of detectors H1 and L1.}
\label{tab:test-gw170817}
\begin{ruledtabular}\begin{tabular}{clllll}
\textbf{Detector}    & $\min{x}_{ij}$ & $\max{x}_{ij}$ & $\overline{x}_{ij}$ & $\sigma_{{x}_{ij}}$ & $\widetilde{x}_{ij}$ \\ 
H1 & 0.273    & 188.188  & 49.205     & 26.095 &  46.069  \\
L1 & 0.200     & 4679.162  & 210.509     & 589.380 &   48.509  \\
\end{tabular}\end{ruledtabular}%
\end{table}

In Table \ref{tab:test-gw170817} we report the minimum, maximum, mean, median, and standard deviation values of the spectrogram bins - with $x_{ij}$ being the absolute value of the bin with indices $ij$ - for the time interval $t_e-4s<t <t_e+4s$ in the L1 and H1 detectors. Whilst the maximum and the mean values of the bins in the spectrogram of the L1 detector are greatly influenced by the glitch, the minimum and median are close to those obtained for the H1 detector. Moreover, the bin values are largely more constrained for the H1 detector - $x_{ij} \in [0.273,\, 188.188]$ - than for the L1 detector - $x_{ij} \in [0.200,\, 4679.162]$. Thus, limiting the permissible range of bin values - or, analogously, limiting the colormap range - before converting the spectrogram to an 8-bit per channel RGB map should allow for better preservation of the details previously obfuscated by the normalization process - assuming that undesirable phenomena, such as harmonic distortion or other artifacts, have little to no impact in the regions outside of the glitch as a result of its presence.

\begin{figure}[!ht]%
    \centering
    \subfloat[\centering $t_e-8s<t <t_e$]{{\includegraphics[width=0.45\linewidth]{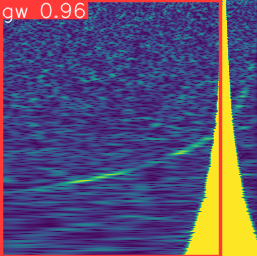} }}%
    \hfill
    \subfloat[\centering $t_e-4s<t <t_e+4s$]{{\includegraphics[width=0.45\linewidth]{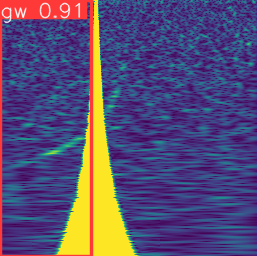} }}%
    \caption{Spectrograms of the GW170817 event data from the LIGO L1 detector limited to $x_{ij} =256$ and respective event detection bounding boxes and confidence value obtained by applying the trained YOLOv5 model}%
    \label{fig:l1v3-scl}%
\end{figure}

To test this hypothesis, an upper limit of 256 was enforced for the absolute values of $x_{ij}$. By introducing this modification, it is possible to obtain much more similar spectrograms - visually - to those obtained for the H1 detector - cf. Figure \ref{fig:l1v3-scl}. As a result, the trained model was able to successfully identify the event in the same two frames as previously, despite the presence of the glitch. No false positives were detected.

Therefore, despite the inability of the model to detect the event in the raw data in the presence of the glitch, this shortcoming was easily solved by applying a simple pre-processing step. While not being the scope of this work, more robust pre-processing pipelines can be idealized to guarantee the best detection conditions for the model. It should also be noted that, despite being trained with synthetic noise based on the PSD of the H1 detector, the model was successful in detecting the event with real detector noise. Moreover, no specific training was performed with samples based on the noise profile of a specific detector, thus proving the vast generalization ability of this approach.

\subsection{Detecting the GW190425 event}

As our final test we consider the data of the GW190425 event  \cite{https://doi.org/10.48550/arxiv.2108.01045} from the LIGO Livingston (L1) observatory. The analysis of this CBC event yielded mass parameters compatible with the individual binary components being neutron stars. The detection SNR of GW190425 is 12.9 \cite{Abbott_2020}, which is compatible with the model proposed in this work.

\begin{figure}[!ht]%
    \centering
    \subfloat[\centering $t_e-8s<t <t_e$]{{\includegraphics[width=0.45\linewidth]{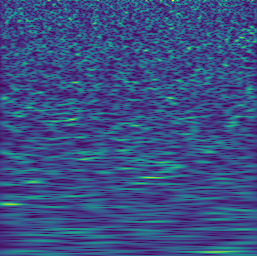} }}%
    \hfill
    \subfloat[\centering $t_e-4s<t <t_e+4s$]{{\includegraphics[width=0.45\linewidth]{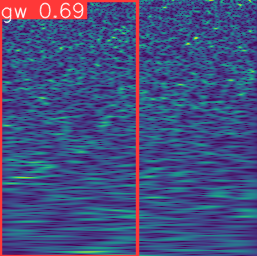} }}%
    \caption{Spectrograms of the GW190425 event data from the LIGO L1 detector near the event time and successful detection with its respective bounding box and confidence value obtained by applying the trained YOLOv5 model}%
    \label{fig:gw1900425_l1v3}%
\end{figure}

\begin{figure}[!ht]%
    \centering
    \subfloat[\centering $t_e-\SI{196}{s}<t <t_e-\SI{188}{s}$]{{\includegraphics[width=0.45\linewidth]{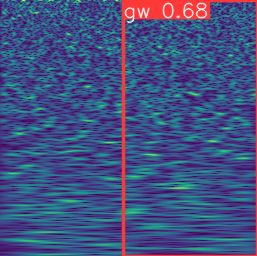} }}%
    \hfill
    \subfloat[\centering $t_e-\SI{180}{s}<t <t_e-\SI{172}{s}$]{{\includegraphics[width=0.45\linewidth]{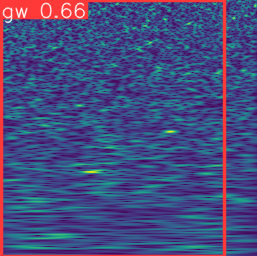} }}%
    \newline
    \subfloat[\centering $t_e-\SI{84}{s}<t <t_e-\SI{76}{s}$]{{\includegraphics[width=0.45\linewidth]{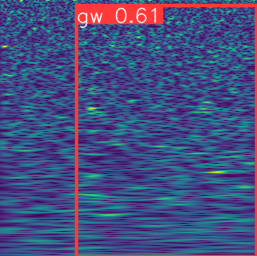} }}%
    \hfill
    \subfloat[\centering $t_e+\SI{12}{s}<t <t_e+\SI{20}{s}$]{{\includegraphics[width=0.45\linewidth]{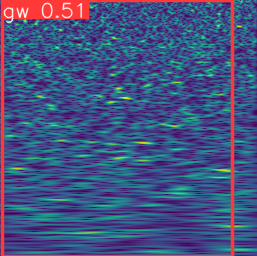} }}%
     \newline
    \subfloat[\centering $t_e+\SI{68}{s}<t <t_e+\SI{76}{s}$]{{\includegraphics[width=0.45\linewidth]{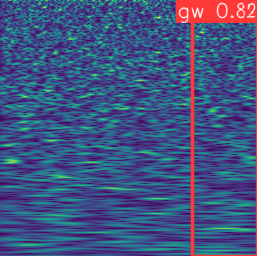} }}%
    \hfill
    \subfloat[\centering $t_e+\SI{76}{s}<t <t_e+\SI{84}{s}$]{{\includegraphics[width=0.45\linewidth]{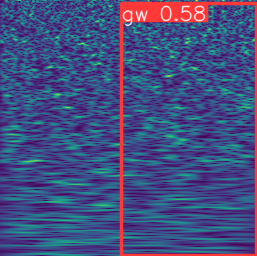} }}%
     \newline
    \subfloat[\centering $t_e+\SI{192}{s}<t <t_e+\SI{200}{s}$]{{\includegraphics[width=0.45\linewidth]{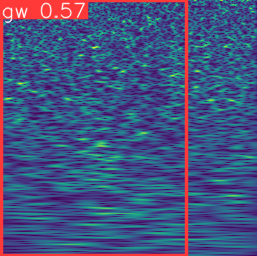} }}%
    \caption{Spectrograms of the GW190425 event data from the LIGO L1 detector containing false-positive detections with their respective bounding boxes and confidence values obtained by applying the trained YOLOv5 model}%
    \label{fig:gw1900425_l1v3_false}%
\end{figure}

Similar to the previous test, we consider a window of 602 seconds centered at the event. After removing $\SI{2}{s}$ of corrupted samples - resulting from the whitening filter - and applying a window with a duration of $\SI{8}{s}$, and a stride of $\si{4}{s}$, we obtain a total of 150 samples. Running the inference pipeline with an IoU threshold of $0.45$ and a confidence threshold of $0.90$ results in no successful detections in the interval considered. However, by setting the confidence threshold to 0.50 we obtain a successful detection of the event in the $t_e-4s<t <t_e+4s$ window with a confidence of 0.69 - see Figure \ref{fig:gw1900425_l1v3}. As expected, the reduction of the confidence threshold led to the increase of false-positive events, in a total of 7 false detections - Figure \ref{fig:gw1900425_l1v3_false}.

\section{Conclusions} \label{sec:conclusions}

In this work we have presented a  gravitational-wave detection approach for CBC sources based on the YOLOv5 model, a general object-detection implementation usually applied in computer vision tasks. Despite this work has been presented as a proof-of-concept of the application of such models for CBC GW detection, some real-world test cases have also been discussed. Those have served to verify the model's ability to correctly operate in actual GW events.

Firstly, we have described the generation pipeline used for creating synthetic CBC spectrogram datasets. Whilst not the main focus of this work, this step is of the utmost importance for obtaining appreciable performance with any machine learning model. Some aspects of this pipeline and of the resulting samples have not been addressed methodically in this work; a few examples include the effects of image resolution, the limited bit-depth of 8-bit RGB images, and the selection of the colormap and the transform method used for generating the spectrogram images. As these may greatly influence the performance of the model, a more thorough analysis of the correctness and possible improvements of this pipeline should be considered in future research.

It should be noted that in this work no real samples - of both noise and CBC GW strains - were used. Whilst it is improbable that a large enough dataset of real CBC strains compatible with machine learning tasks can be achieved in the foreseeable future, detector noise is widely available and should be considered for further development. The inclusion of real detector noise should allow for improvement in the generalization abilities of the model, as it introduces a wide variety of phenomena not contemplated in synthetic noise samples generated from PSD profiles. 

Having generated the dataset, the training of the model was performed, returning an appreciable evolution of both the validation metrics and loss values - cf. Appendix \ref{appendix:metrics}. We have found that the value of the optimization metric - mAP - reached in the best epoch of the training is considerably greater than that usually achieved in usual object detection benchmark tasks, even with models from the YOLO family \cite{Sultana_2020, 8627998}. However, it should be noted that we are currently only dealing with a single class dataset.

Finally, various tests have been applied to the trained model to verify the quality of the training and its capabilities in detecting both artificial and real GW events. Despite the tests with synthetic data do not provide a definitive analysis of the quality of the training, they hinted that the model was able to successfully learn the features of CBC gravitational 
-wave events in spectrograms and that it should be capable of performing inference in real detector data. This has been further verified by applying the model in the GW170817 event yielding a successful detection in both the H1 and the L1 detectors. The detection of the GW190425 event with the YOLOv5 model has been, however, less successful. This attests to the performance degradation phenomenon resulting from the decrease in SNR observed in the tests, as GW190425 has a significantly lower SNR than the GW170817 signal.

The model's ability to perform real-time detection on a single detector with reasonable confidence, as well as its intrinsic capability of precisely providing bounds for the time of the events, make it an interesting approach for both auxiliary first-stage detection alarm pipelines and the integration in more complex pipelines - e.g. for real-time physical parameter estimation. 
Moreover, one of the foreseeable advantages of this model is its ability to perform multi-class detection and to provide multiple detections in a single sample. As such, YOLO-based pipelines for simultaneous multi-object detection - such as all types of transient CBC events (BNS, BBH, and black hole - neutron star binaries) and glitches - seem viable and can be easily developed using this model.

\section*{Acknowledgements}

JA acknowledges support by the project IMFire - Intelligent Management of Wildfires, ref. PCIF / SSI / 0151/2018, and was fully funded by national funds through the Ministry of Science, Technology, and Higher Education. %
FFF is supported by the FCT project PTDC/FIS-PAR/31000/2017 and by the Center for Research and Development in Mathematics and Applications (CIDMA) through FCT, references UIDB/04106/2020 and UIDP/04106/2020. %
MF and CP acknowledges partial  support  by national funds from FCT (Fundação para a Ciência e a Tecnologia, I.P, Portugal) under the Projects No. UID/\-FIS/\-04564/\-2019, No. UIDP/\-04564/\-2020, No. UIDB/\-04564/\-2020, and No. POCI-01-0145-FEDER-029912 with financial support from Science, Technology and Innovation, in its FEDER component, and by the FCT/MCTES budget through national funds (OE). %
JAF acknowledges support from the Spanish Agencia Estatal de  Investigaci\'on  (PGC2018-095984-B-I00) and  from  the Generalitat Valenciana (PROMETEO/2019/071). %
AO acknowledges support from national funds from FCT, under the project CERN/FIS- PAR/0029/2019.%
The authors acknowledge the Laboratory for Advanced Computing at the University of Coimbra (http://www.uc.pt/lca) for providing access to the HPC computing resource Navigator,
Minho Advanced Computing Center (MACC) for providing HPC resources that have contributed to the research results reported within this paper, the Portuguese National Network for Advanced Computing for the grant CPCA/A1-428291-2021. %
Finally, the authors gratefully acknowledge the computer resources at Artemisa, funded by the European Union ERDF and Comunitat Valenciana as well as the technical support provided by the Instituto de Física Corpuscular, IFIC (CSIC-UV). %

\bibliographystyle{apsrev}
\bibliography{references.bib}

\appendix

\section{Training metrics and loss value}
\label{appendix:metrics} 

\begin{figure*}[!ht]%
    \centering
    \subfloat[\centering mAP\textsubscript{0.50}]{
        \includegraphics[width=0.35\linewidth]{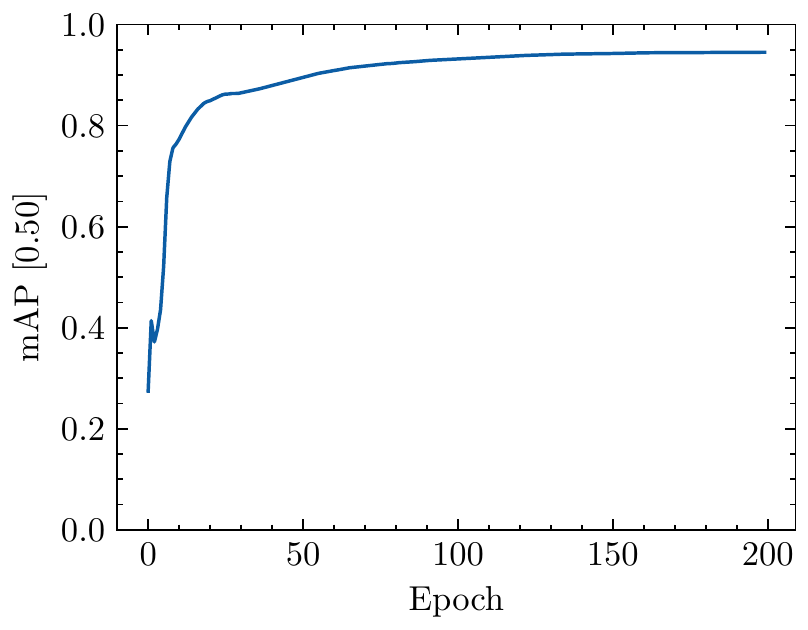} 
    }%
    \quad \quad \quad
    \subfloat[\centering mAP\textsubscript{0.50:0.95}]{
        \includegraphics[width=0.35\linewidth]{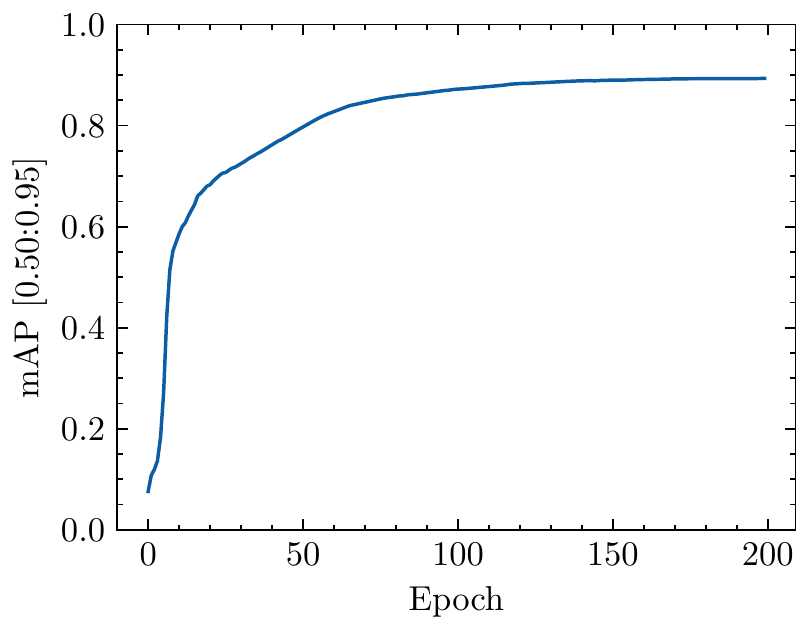}
    }\\
    \subfloat[\centering Precision]{
        \includegraphics[width=0.35\linewidth]{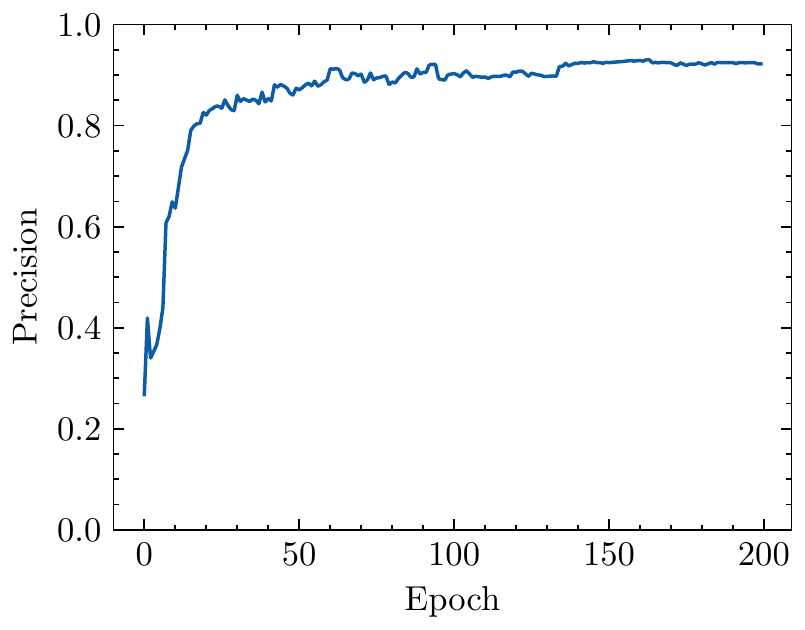} 
    }%
    \quad \quad \quad
    \subfloat[\centering Recall]{
        \includegraphics[width=0.35\linewidth]{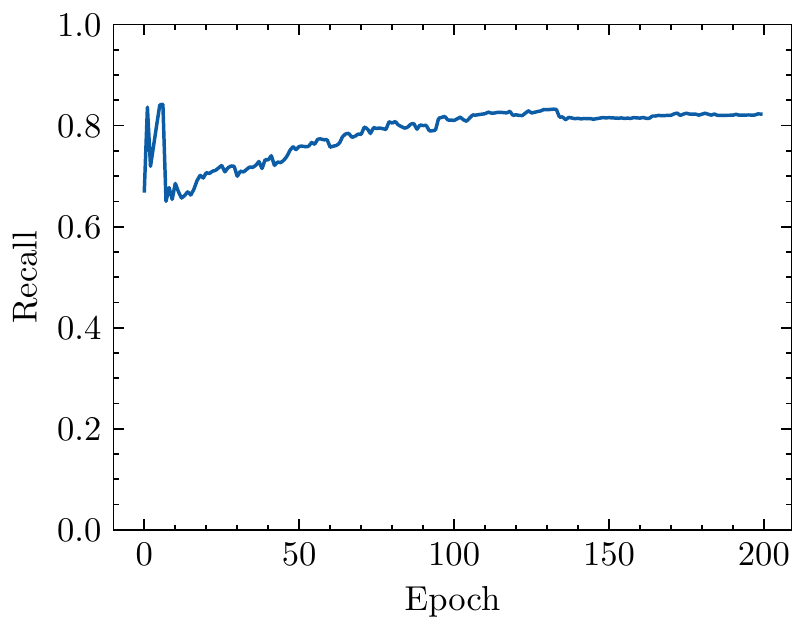} 
    }\\
    \centering
    \subfloat[\centering Object Loss (Training)]{
        \includegraphics[width=0.35\linewidth]{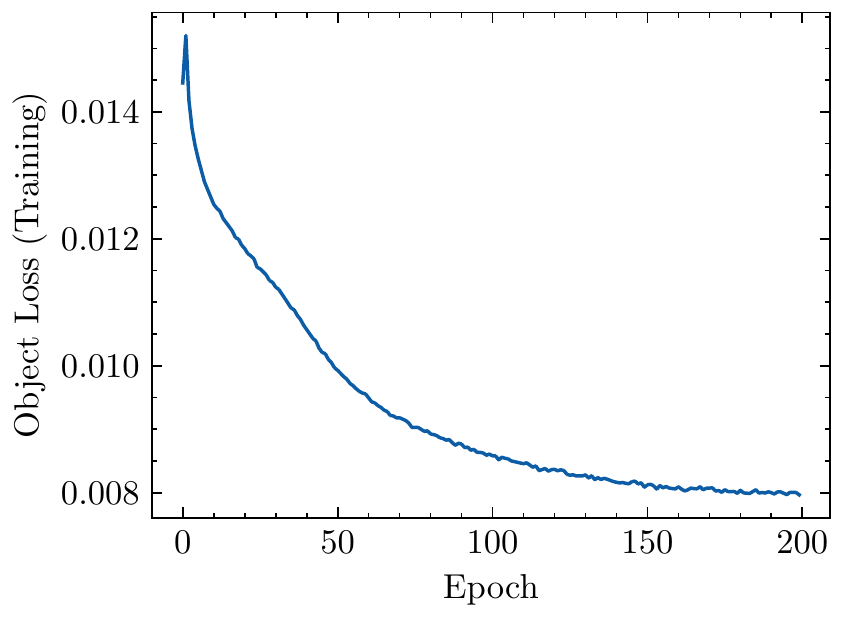} 
    }%
    \quad \quad \quad
    \subfloat[\centering Object Loss (Validation)]{
        \includegraphics[width=0.35\linewidth]{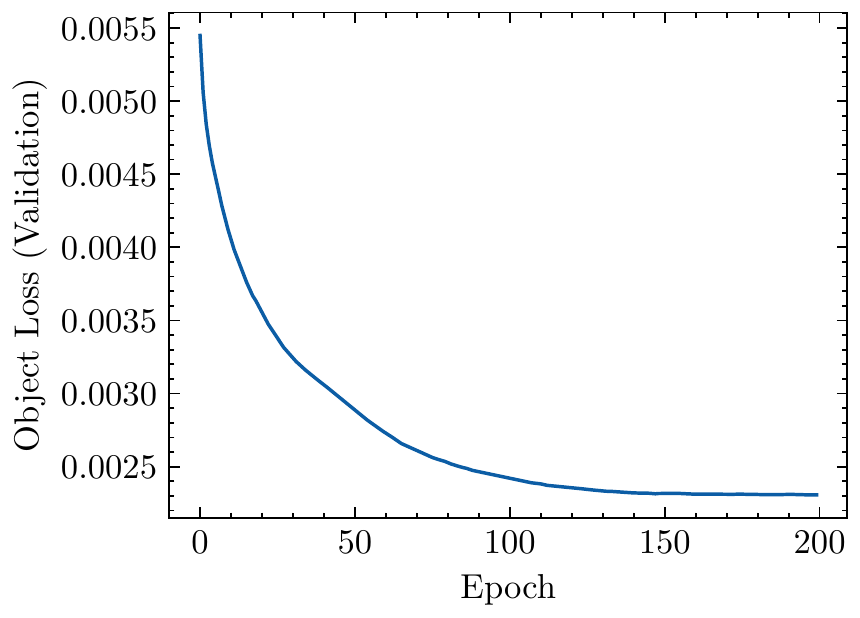} 
    }\\
    \subfloat[\centering Box Loss (Training)]{
        \includegraphics[width=0.35\linewidth]{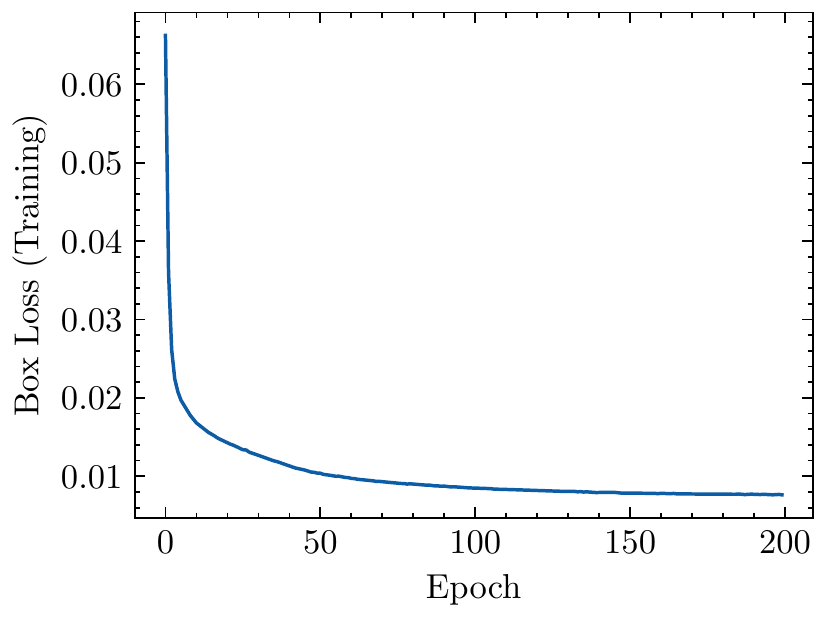} 
    }%
    \quad \quad \quad
    \subfloat[\centering Box Loss (Validation)]{
        \includegraphics[width=0.35\linewidth]{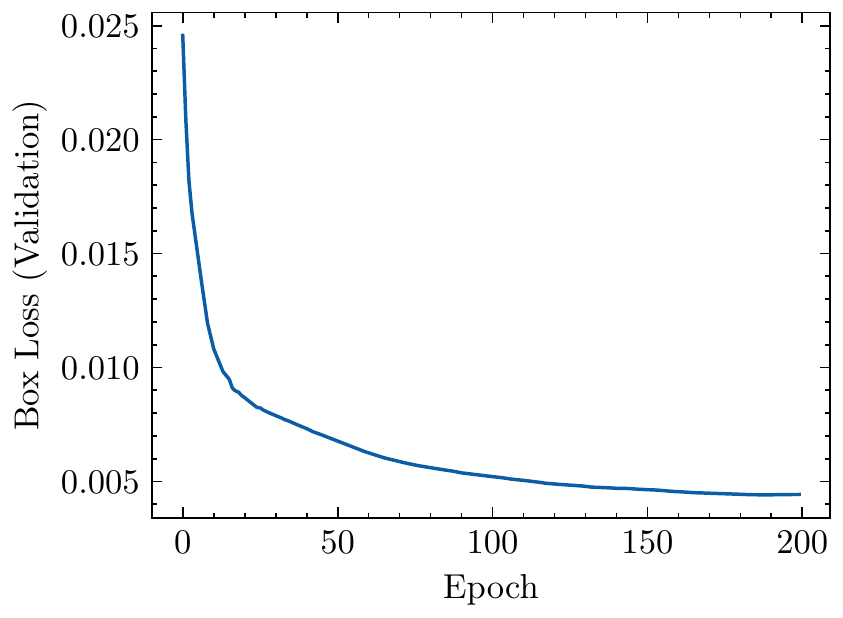} 
    }\\
    \caption{Evolution of training metrics and loss as a function of the epochs for the model training process, namely (a) Mean Average Precision [0.50], (b) Mean Average Precision [0.50:0.95], (c) Precision, (d) Recall, (e) Object Loss (training), (f) Object Loss (validation), (g) Box Loss (training), and (h) Box Loss (validation).} %
    \label{fig:train_metrics_losses}%
\end{figure*}

The training of the proposed model consisted of a total of 200 epochs. The evolution of various metrics and both the bounding box and the object losses is shown in Figure~\ref{fig:train_metrics_losses}.

\end{document}